%% file: draft.tex
\documentclass[aps,pre,twocolumn,showpacs,amsmath,amssymb]{revtex4-1}
\usepackage[pdfusetitle,pdfauthor={Navdeep Rana}]{hyperref}
\usepackage[dvipsnames]{xcolor}
\pagecolor{white}
\usepackage[final]{graphicx}
\usepackage{bm}
\usepackage{enumerate}
\usepackage{enumitem}
\usepackage[normalem]{ulem}
\usepackage[capitalize]{cleveref}
\crefname{equation}{Eq.}{}
\crefrangeformat{equation}{Eqs.~(#3#1#4)-(#5#2#6)}

\input{abbreviations.tex}

\newcommand{\Sa}{\bm{\Sigma}^{a}}
\newcommand{\Sr}{\bm{\Sigma}^{r}}

\newcommand{\REM}[1]{{}}

\begin{document}
\title{Defect turbulence in a dense suspension of polar, active swimmers}
\author{Navdeep Rana}
\affiliation{Max Planck Institute for Dynamics and Self-Organization (MPIDS), D-37077 G\"ottingen, Germany}
\author{Rayan Chatterjee}
\affiliation{Stanford Medicine, Stanford University, USA}
\author{Sunghan Ro}
\affiliation{Department of Physics, Technion-Israel Institute of Technology, Haifa 3200003, Israel}
\author{Dov Levine}
\affiliation{Department of Physics, Technion-Israel Institute of Technology, Haifa 3200003, Israel}
\author{Sriram Ramaswamy}
\affiliation{Department of Physics, Indian Institute of Science, Bengaluru 560 012 India}
\author{Prasad Perlekar}
\email{perlekar@tifrh.res.in}
\affiliation{Tata Institute of Fundamental Research, Hyderabad, India}

\begin{abstract}
    We study the effects of inertia in dense suspensions of polar swimmers. The hydrodynamic velocity field and the
    polar order parameter field describe the dynamics of the suspension. We show that a dimensionless parameter $R$
    (ratio of the swimmer self-advection speed to the active stress invasion speed \cite{chatterjee2021}) controls the
    stability of an ordered swimmer suspension. For $R$ smaller than a threshold $R_1$, perturbations grow at a rate
    proportional to their wave number $q$. Beyond $R_1$, we show that the growth rate is $\mathcal{O}(q^2)$ until a
    second threshold $R=R_2$ is reached.  The suspension is stable for  $R>R_2$. We perform direct numerical simulations
    to investigate the steady state properties and observe defect turbulence for $R<R_2$. An investigation of the
    spatial organisation of defects unravels a hidden transition: for small $R\approx 0$ defects are uniformly
    distributed and cluster as $R\to R_1$. Beyond $R_1$, clustering saturates and defects are arranged in nearly
    string-like structures.
\end{abstract}

\maketitle

\section{Introduction}

The instability \cite{simha2002, chatterjee2021} of a viscous suspension of motile organisms leads to an abundance of
ordered and chaotic states
\cite{marchetti2013,ramaswamy2010,wensink2012,dombrowski2004,giomi2015,lauga2016,lauga2012,lauga2009}, of which
\textit{active turbulence} \cite{alert2022} has attracted much attention. The majority of studies have investigated this
iconic phenomenon in \textit{nematic} systems, in which the local structure and characteristic defects are associated
with uniaxial fore-aft symmetric order. In a recent work \cite{chatterjee2021}, we established the stabilizing role of
fluid inertia in the transition to active turbulence in \textit{polar} active suspensions, in which the local order is
vectorial. We discovered that the steady-state behavior of the suspension changed across two thresholds $R_1$ and $R_2$
in the dimensionless ratio $R \equiv \rho v_0^2/\sigma$ of inertia and activity, where $\rho$ is the mass density of the
suspension, $v_0$ is the self-propulsion speed, and $\sigma$ the mean active stress. We found defect turbulence for
$R<R_1$, a fluctuating but ordered phase-turbulent steady state for $R_1<R<R_2$, and a quiescent ordered state for $R >
R_2$. The analysis of \cite{chatterjee2021} considered only the suspension velocity and polar order-parameter fields
$\bm{u}$ and $\bm{p}$, implicitly treating the concentration of active particles as ``fast'', as would happen, for
example, if birth and death led to a nonzero relaxation rate at zero wavenumber. 

Highly concentrated systems present another scenario where we can ignore fluctuations in the number density of suspended
particles, albeit in a radically different manner. In such a limit, it is reasonable to make the approximation that
concentration fluctuations are prohibited, which means that the solute velocity field or, equivalently, the polar order
parameter $\bm{p}$, is solenoidal: $\nabla \cdot \bm{p} = 0$. We will term such systems as simply ``dense''. See
\cite{chen2016} for the implementation of such a condition in dry active matter. 

In this paper, we study stability and turbulence in a dense suspension of polar active particles (DSPAP).
\cref{sec:equations} describes the equations of motion and in \cref{sec:lsa}, we conduct a linear stability analysis of
an ordered suspension with respect to perturbations with wavenumber $q$. We show that for extensile suspensions, the
same dimensionless parameter $R$ as in \cite{chatterjee2021} governs the linear stability of the suspension. For $R <
R_1 \equiv 1 + \lambda$, where $\lambda$ is the flow-alignment parameter, we find an inviscid instability where the most
unstable pure bend modes grow at a rate $\propto q$. For $R_1 < R < R_2$, we find that the pure bend perturbations grow
at a rate $\propto q^2$. Finally for $R > R_2$, orientational order is linearly stable. These instability mechanisms are
identical to those in our earlier report \cite{chatterjee2021}. Further, dense suspensions of \textit{contractile}
swimmers are linearly stable. Next, in \cref{sec:numerical}, we study the steady-state properties of the suspension
using numerical simulations. We show the presence of vortex-defect turbulence for $R < R_2$. Correlation length  grows
as we increase $R$ and appears to diverge at $R \approx R_2$. Interestingly we do not observe a phase-turbulent regime
\cite{chatterjee2021}. Instead, a detailed analysis of the spatial organization of defects using correlation dimension
$d_2$ reveals a novel morphological transition. We find that defects are uniformly distributed ($d_2 \approx 2$) for $R
\to 0$ (Stokesian suspension) and start to cluster on increasing $R$. Maximum clustering ($d_2 \approx 1$) is attained
around $R = R_1$, and we observe that defects organize into stringy patterns. No further changes in defect organization
are observed on increasing $R$ beyond $R_1$.

\section{\label{sec:equations} Equations of motion}
We use a hydrodynamic formulation to study the dense suspension of polar active particles (DSPAP). In the incompressible
limit (uniform suspension density and active particle concentration), the dynamics of the hydrodynamic velocity field
$\vel\postime$ and the orientation order parameter $\pol\postime$ is described by the following equations
\cite{simha2002,cates2018,lenz2008,marchetti2013,chatterjee2021,kruse2005}:
\begin{equation} \label{eq:dspp} \begin{aligned}
    & \rho\left(\dti \vel + \vel \cdot \nabla \vel \right) =
    - \nabla P + \mu \nabla^2 \vel + \nabla\cdot(\Sa + \Sr), \\
    & \dti \pol + \left(\vel+v_0\pol\right)\cdot\nabla\pol =
    - \nabla \Pi + \left(\lambda\bm{S} + \bm{\Omega}\right)\cdot\pol + \Gamma\res,\\
    & \nabla \cdot \vel=0, ~{\rm and}~\nabla \cdot \pol=0.
\end{aligned} \end{equation}
Here, $\rho$ is the suspension density, $\mu$ is the fluid viscosity, $\vo$ is the self-advection speed of the swimmers,
$\lambda$ is the flow alignment parameter,  $P$ is the hydrodynamic pressure that enforces incompressibility of the
velocity field, $\Pi$ is the pressure-like term that enforces incompressibility of the order parameter field $\pol$ 
imposed by the constant concentration approximation, $\bm{S}\equiv (\nabla\vel + \nabla\vel^T)/2$ and $\bm{\Omega}\equiv
(\nabla\vel - \nabla\vel^T)/2$ are the symmetric and anti-symmetric parts of the velocity gradient tensor $\nabla\vel$.
$\Sa \equiv -\so \pol\pol$ is the leading order apolar intrinsic stress associated with the swimming activity where the
force-dipole density $\so > 0 \left(< 0\right)$ for extensile (contractile) swimmers \cite{simha2002}, $\Sr \equiv
-\lp\res\pol - \lm\pol\res$ is the reversible thermodynamic stress \cite{kruse2005} with $\lambda_{\pm} = (\lambda \pm
1)/2$, $\res =-{\delta F}/{\delta \pol}$ is the molecular field conjugate to $\pol$ with the free-energy functional
\begin{equation} \begin{aligned}
    F = \int \dd^{d}r \left[\frac{K}{2} |\nabla \pol |^2
    +\frac{1}{4}\left(\pol\cdot\pol-1\right)^2\right],
    \nonumber
\end{aligned} \end{equation}
that favors an aligned order parameter state with unit magnitude, and  $\Gamma$ is the rotational mobility for the
relaxation of the order parameter field to the uniform ordered state prescribed by the free energy dynamics. For
simplicity, we choose a single Frank constant $K$, which penalizes the gradients in $\pol$ \cite{gennes1993}. We are
primarily interested in the interplay of the self-propulsion speed $v_0$ and the leading order apolar active stress,
hence similar to our earlier study \cite{chatterjee2021}, we have ignored the contribution from higher-order polar
gradient terms in $\Sa$.

\begin{figure}
    \centering
    \includegraphics[width=\linewidth]{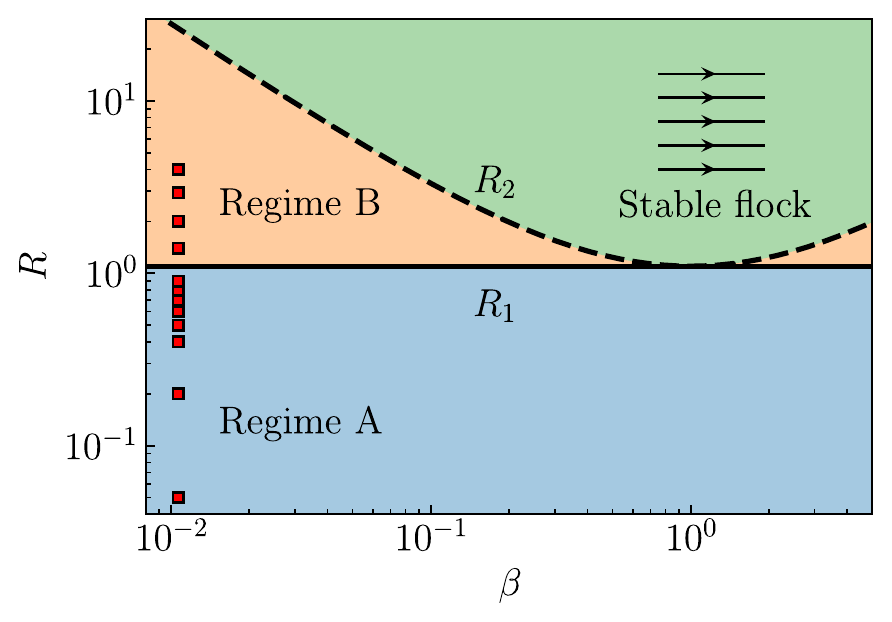}
    \caption{ \label{fig:phase-diagram}
        $R-\beta$ phase diagram highlighting different stability regimes for incompressible polar active suspensions. In
        both the unstable regimes A and B, we observe defect turbulence (see \cref{sec:numerical}). Red squares mark
        the simulations on the $R-\beta$ plane with $\beta = 10^{-2}$, $R_{1} = 1.1$, and $R_{2} = 28$. See
        \cref{tab:parameters} for the rest of parameters.
    }
\end{figure}

\section{\label{sec:lsa} Linear stability Analysis}
We analyse the stability of the ordered state $\left(\vel,\pol\right) = (0,\icap)$ to small monochromatic perturbations
of the form $\left(\bm{\delta u},\bm{\delta p}\right) = \left(\widehat{\vel},\widehat{\pol}\right)
e^{i(\bm{q}\cdot\bm{x} -\omega t)}$, where $\bm{q}$ is the perturbation wavevector and $\omega$ is the frequency. As
dense contractile suspension are linearly stable, here, we focus our study on extensile suspension. In what follows, we
discuss the results for pure-bend perturbations, the most unstable modes for extensile systems
\cite{ramaswamy2010,chatterjee2021}. For a detailed discussion of the linear stability analysis including the stability
of twist-bend and splay-bend modes, we refer the reader to \cref{sec:appendix}. A large wavelength (small $q$) expansion
up to $\mathcal{O}(q^2)$ yields the following dispersion relation for the pure-bend modes:
\begin{equation}\label{eq:smallq} \begin{aligned}
    2\omega_{\pm} &= \vo \left(1 \pm \sqrt{1-R_{1}/R}\right)q \\
                  &+ i\frac{\mu}{\rho} \left(\pm\frac{(1-\beta)}{\sqrt{1 - R_{1}/R}}-(1+\beta) \right)q^2,
\end{aligned} \end{equation}
where we have defined the dimensionless numbers $R \equiv \rho \vo^2/2\so$, $R_{1} \equiv 1+\lambda$ and $\beta = \Gamma
K \rho/\mu$. In \cref{fig:phase-diagram}, we show a qualitative phase diagram highlighting the three different stability
regimes. In regime A, $R < R_1$ and irrespective of the value of $\beta$, pure-bend modes are unstable with a growth
rate $\imw \propto q$. In regime B, $R_1 < R < R_2$ where $R_{2} \equiv R_1\left( 1+\beta \right)^2/4\beta$, pure-bend
modes grow at a rate $\imw \propto q^{2}$. Finally, in regime C, when $R > R_2$ the ordered state is stable.

\section{\label{sec:numerical} Direct Numerical Simulations}

\begin{figure*}
    \centering
    \includegraphics[width=\linewidth]{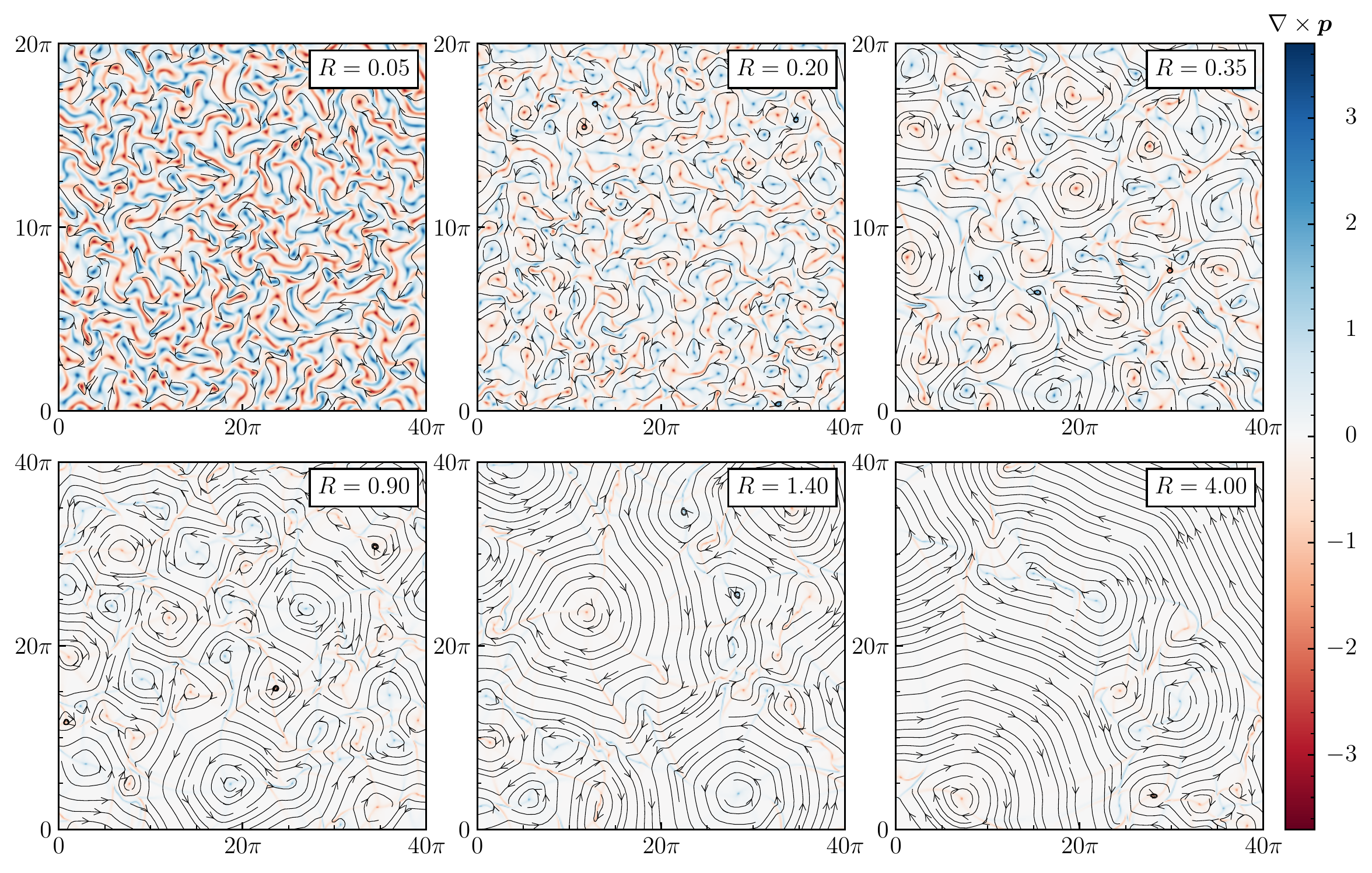}
    \caption{\label{fig:snapshot}
        Order parameter streamlines superimposed over pseudocolor plot of $\nabla\times\pol$ highlighting vortices at
        different values of R. For $R \geq 0.9$, we only show a subdomain of size $(40\pi)^2$. Inter-defect separation
        grows with increasing $R$. For $R < R_1 = 1.1$, we find that the defects are uniformly distributed over the simulation
        domain, but form clusters for $R > R_1$ (See also \cref{fig:clustering}).
    }
\end{figure*}

We perform direct numerical simulations of \eqref{eq:dspp} on a square domain of area $L^{2}$ discretized with
$N^2$ equispaced collocation points. Similar to \cite{chatterjee2021}, we use a pseudo-spectral method for spatial
integration of the velocity field and a fourth-order finite difference method for the order parameter. For time marching
we use a second-order Adams-Bashforth scheme \cite{cox2002}, and the incompressibility condition in the order-parameter
field is implemented using an operator-splitting method \cite{canuto1988}. We undertake high-resolution numerical
studies at various values of $R$ and fixed $\beta = 10^{-2}$ to characterize the turbulent states arising from the
instabilities of the ordered state. We initialize our simulations with a perturbed ordered state $\vel({\bm x},0) =
{\bm 0} + A \sum_{i=1}^{10} \cos\left(q_i x\right) \jcap$ and $\pol({\bm x},0) = \icap + B \sum_{i=1}^{10} \cos\left(q_i
x\right)\jcap$,
where $ q_i = i (2\pi/{L})$, and choose $A = B = 10^{-3}$. For $R<R_2$, the perturbations destabilise the flow and a
defect turbulence state is achieved. \cref{tab:parameters} summarizes all our simulation parameters. In what follows, we
discuss the statistical properties of defect turbulence with varying $R$.

\begin{table}[!b]
    \centering
    \begin{tabular}{@{\extracolsep{\fill}} c c c c }
        \hline
        $L$ & $N$ & $R \equiv \Rdef$ \\
        \hline
        \hline
        $10\pi$ & $1024$ & $0.05$ \\
        $40\pi$ & $2048$ & $0.2-0.4^{\star}$ \\
        $80\pi$ & $4096$ & $0.5$ \\
        $160\pi$ & $4096$ & $0.6-0.9^{\dagger}$ \\
        $160\pi$ & $4096$ & $1.4,~2,~3,~4$ \\
        \hline
        \hline
    \end{tabular}
    \caption{\label{tab:parameters}
        Parameters used in our simulations: $\rho=1$, $\lambda=0.1$, $\mu=0.1$, $K=10^{-3}$, $\vo=3.16\times10^{-2}$, and
        $\Gamma = 1$ are kept fixed for all runs. This sets $\beta = 10^{-2}$, $R_{1} = 1.1$, and $R_{2} = 28$. We vary
        $R$ by varying $\sigma_0$. Superscript ${\star}$ ($\dagger$) indicate increments of $0.05$ ($0.1$).
    }
\end{table}

\begin{figure}
    \includegraphics[width=\linewidth]{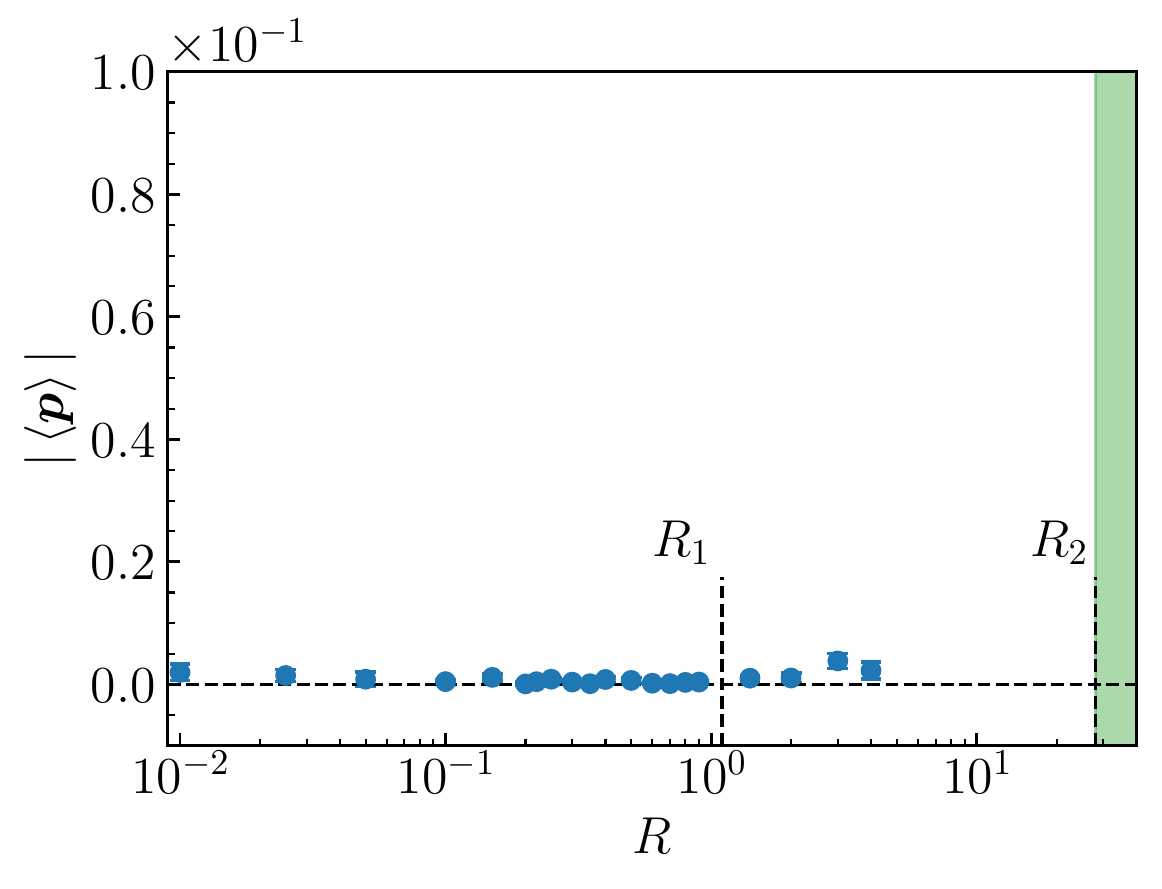}
    \caption{\label{fig:average-order}
        Average order $|\left<\bm{p}\right>|$ in the steady state at different $R$. For a disordered state
        $|\left<\bm{p}\right>|=0$ and for perfect alignment, $|\left<\bm{p}\right>|=1$. We do not observe any order in
        regime A or regime B. Ordered state is stable to perturbations in the green shaded region ($R > R_2$). The
        error bars are smaller than the markers.
    }
\end{figure}

\begin{figure*}
    \centering
    \includegraphics[width=\linewidth]{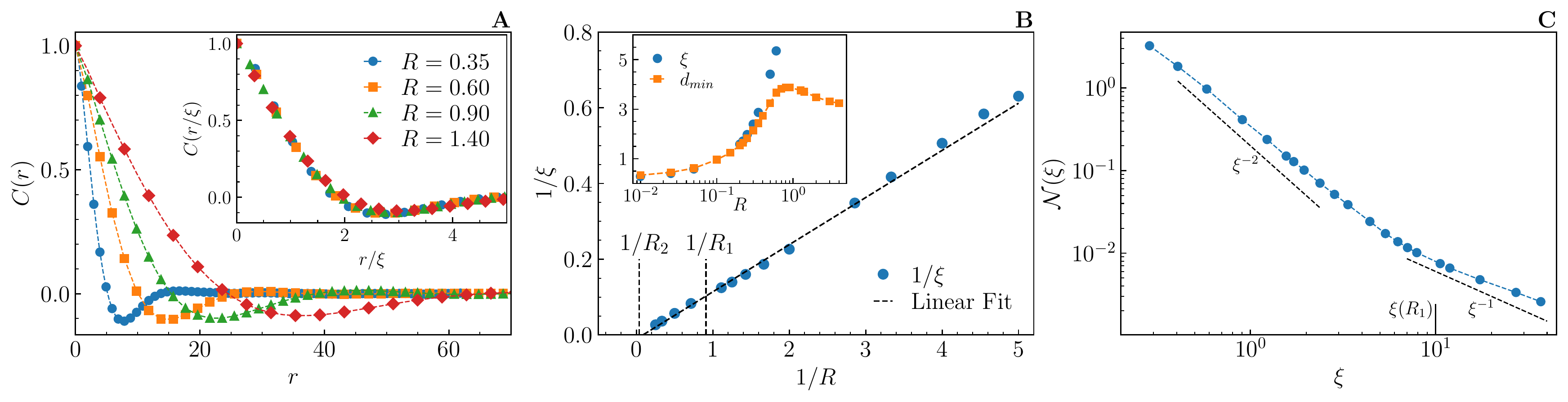}
    \caption{ \label{fig:correlations}
        (A) Steady state correlation function $C(r)$ for different values of R. (Inset) Correlation function collapse
        onto a single curve when distance is scaled by the correlation length $\xi$. (B) Plot of the inverse correlation
        length $1/\xi$ versus $1/R$. From the intercept of the linear fit on the $1/R$ axis, we conclude that $\xi$
        diverges around $R \approx R_2$. Inset: Comparison between the correlation length $\xi$ and average
        nearest-neighbour distance $d_{min}$. For small $R$, $\xi$ and $d_{min}$ are identical. For large $R$, $d_{min}$
        saturates and $\xi$ diverges. (C) Defect density $\mathcal{N}(\xi)$ as a function of the correlation length
        $\xi$. We observe two distinct scaling regimes for $R < R_1 (> R_1)$. Vertical black line marks the correlation
        length $\xi(R=R_1)$ computed from the linear fit in (B).
    }
    \includegraphics[width=\linewidth]{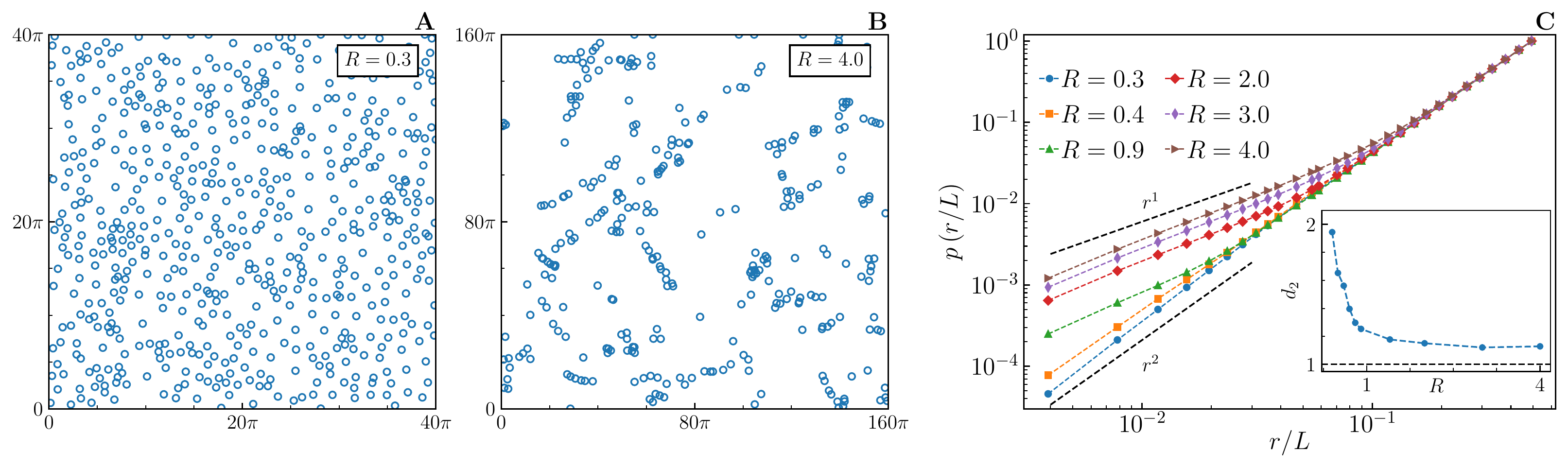}
    \caption{ \label{fig:clustering}
        (A,B) Scatter plot of the vortex cores for $R=0.3$ and $R=4.0$ respectively. For $R=0.3$ and $R \ll R_1$ in
        general, defects are uniformly distributed over the simulation domain. For $R \gtrsim R_1$, we observe
        clustering of defects on one-dimensional string like structures. (C) Plot of the cumulative radial distribution
        function $p(r)$ for different values of $R$. At smaller distances, we observe $p(r) \sim r^{d_2}$, with $d_2
        \sim 2$ for $R \ll R_1$ and $d_2 \sim 1$ for $R \geq R_1$ $p(r)\sim r^1$. At distances larger than typical
        cluster size, $p(r) \sim r^2$ shows the same scaling for all $R$. Note that we have rescaled each $p(r)$ curve
        by its maximum value to highlight different scaling regimes. Inset: Variation of the correlation dimension
        $d_{2}$ vs $R$ obtained from least-square fit on $p(r)$ for small $r$.
    }
\end{figure*}

\subsection{Defect turbulence}
In \cref{fig:snapshot} we show the streamlines of the order parameter field $\pol$ in the statistically steady state at
different values of R. In both regimes A and B, the order parameter field shows vortices and saddles, with
no evidence of global polar order. Asters and spirals are ruled out by the incompressibility constraint. To further
verify that the turbulent states lack global order, we compute the magnitude of the average polar order parameter
$\left|\left< \pol \right>\right|$ in the statistical steady state at different values of $R$, where
$\left<\ldots\right>$ denotes spatio-temporal averaging. Note that for the disordered states, $\left|\left< \pol
\right>\right|=0$, whereas $\left|\left< \pol \right>\right|=1$ for a perfectly aligned state. In \cref{fig:average-order} we
plot $\left|\left< \pol \right>\right|$ with increasing $R$, and, as expected, it is close to zero in both the unstable
regimes.

\subsection{Defect correlations and clustering}
In \cref{fig:correlations}(A), we plot the isotropic correlation function
\begin{equation}\label{eq:correlation}
    C(r) = \frac{\left< \pol\left(\bm{x+r}\right)\cdot\pol\left(\bm{x}\right)\right>}{\left< \pol\left(\bm{x}\right)\cdot\pol(\bm{x}) \right>},
\end{equation}
where $\left<\ldots\right>$ denote ensemble and angle averaging, for different values of $R$. With increasing $R$, the
correlations of the order parameter increase. We fit the functional form $C(r) =
e^{-\left(\frac{r}{\xi}\right)^{\delta}}$ at small $r \ll L$ to extract the correlation length $\xi$ and the exponent
$\delta$ which decreases monotonically from $\delta \approx 1.7$ for $R=0.2$ to $\delta
\rightarrow 1$ for large $R$. The correlation function $C(r)$ collapses onto a unique curve when plotted versus the
scaled separation $r/\xi$ (see \cref{fig:correlations}(A,inset)). In \cref{fig:correlations}(B), we show that the correlation length
$\xi$ increases with $R$, and from the intercept of the linear fit on the $1/R$ axis, it appears to diverge at
$R=R_{2}$. Note that investigating the spatial structure of the order parameter field for $R\to R_{2}$ becomes
numerically unfeasible as $\xi \rightarrow \infty$ and finite-size effects become important.

To quantify the spatial distribution of the defect cores, we begin by identifying the defect coordinates. In
\cref{fig:correlations}(B,inset), we plot the average nearest-neighbor separation $d_{min}$ for different values of $R$. For
$R<R_1$, the correlation length $\xi$ and $d_{min}$ are indistinguishable, indicating that a unique length scale
describes the defect dynamics \footnote{ In the limit $R \rightarrow
    0$, we expect $\xi$ (or $d_{min}$) to saturate to a value proportional to the core-radius of defects $\ell_c \equiv
\sqrt{K}$ (independent of $R$).}. In contrast, for $R>R_1$, correlation length increases, whereas the average nearest
neighbour separation $d_{min}$ saturates. Consequently, in \cref{fig:correlations}(C), the defect number density
$\mathcal{N}(\xi)$ (number of defects per unit area), also scales differently with the correlation length $\xi$ for
$R<R_1$ and $R>R_1$. For $R < R_1$, where a single length scale governs the dynamics, we observe $\mathcal{N}(\xi)\sim
\xi^{-2}$, which indicates that the defects are distributed uniformly over the entire domain
\cite{[{}][{, Appendix VII.}]chandrasekhar1943,hertz1909,rana2020}. In contrast, for $R > R_1$, we find that $\mathcal{N}(\xi) \sim \xi^{-1}$,
indicating clustering of defects.

Indeed, the scatter plots of the defect coordinates in \cref{fig:clustering}(A,B) indicate uniformly distributed defects
for $R < R_1$, whereas they appear clustered for $R>R_1$. We further quantify the clustering by evaluating the
correlation dimension $d_2$ from the defect positions. The correlation dimension $d_2$ is evaluated from the $r\to0$
scaling behaviour of the probability $p(r)$ of finding two defects within a distance $r$
\cite{grassberger1983,mitra2018}. For $R\ll 1$, we find $d_2\sim 2$, and it decreases with increasing R until it
saturates around $d_2\sim 1$ for $R > R_1$ (see \cref{fig:clustering}(C)). Thus we conclude that in dense suspensions,
the cross-over in the $O(q)$ to $O(q^2)$ instability around $R=R_1$ is marked by an intriguing defect clustering
transition.

\begin{figure}
    \centering
    \includegraphics[width=\linewidth]{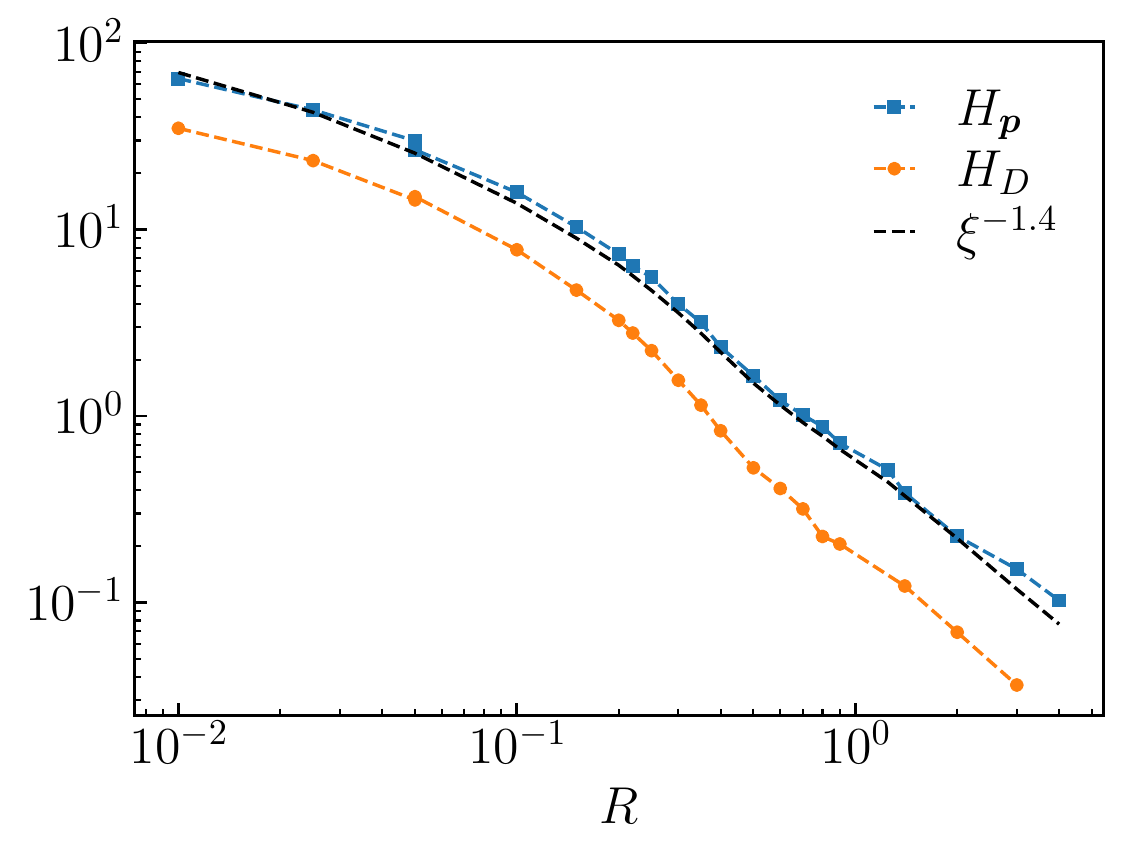}
    \caption{ \label{fig:entropy}
        Plot of the Shannon entropy density $H_{\pol}$ of the polar order parameter and the Shannon entropy density of
        the defect arrangement $H_D$. With increasing $R$, both $H_{\pol}$ and $H_{D}$ decrease and scale roughly as
        $\xi^{-1.4}$ (dashed black guiding line).
    }
\end{figure}

In order to gain further insight on the defect clustering transition, we compute the Shannon entropy density of the
polar order parameter $H_{\pol}$ and the defect arrangement $H_{D}$ using a two-dimensional extension of the pattern
matching method from information theory~\cite{wyner1998, martiniani2019, ro2022}. To find $H_{\bm{p}}$, we apply the
pattern matching method on the discretized orientation field \footnote{$\theta\postime =
\tan^{-1}\left(\frac{p_{y}}{p_{x}}\right)$}. For the entropy density $H_D$, we apply the pattern matching algorithm on a
boolean field which is set to one at the defect locations and zero everywhere else (See \cref{fig:clustering}). In
\cref{fig:entropy}, we plot $H_{\bm{p}}$ and $H_{D}$ versus $R$. As the defect position field contains less information
in comparison to the order parameter field, $H_D$ is smaller than $H_{\bm p}$ for all $R$. Both entropy densities
decrease as we increase $R$, and scale inversely with the correlation length roughly as $H_{D}, H_{\pol} \approx
\xi^{-1.4}$. However, we could not capture a pronounced change in the trend of $H_{\pol}$ or $H_D$ that corresponds to
the clustering transition around $R \simeq R_1$.

\subsection{Energy Spectrum}
\begin{figure}
    \centering
    \includegraphics[width=\linewidth]{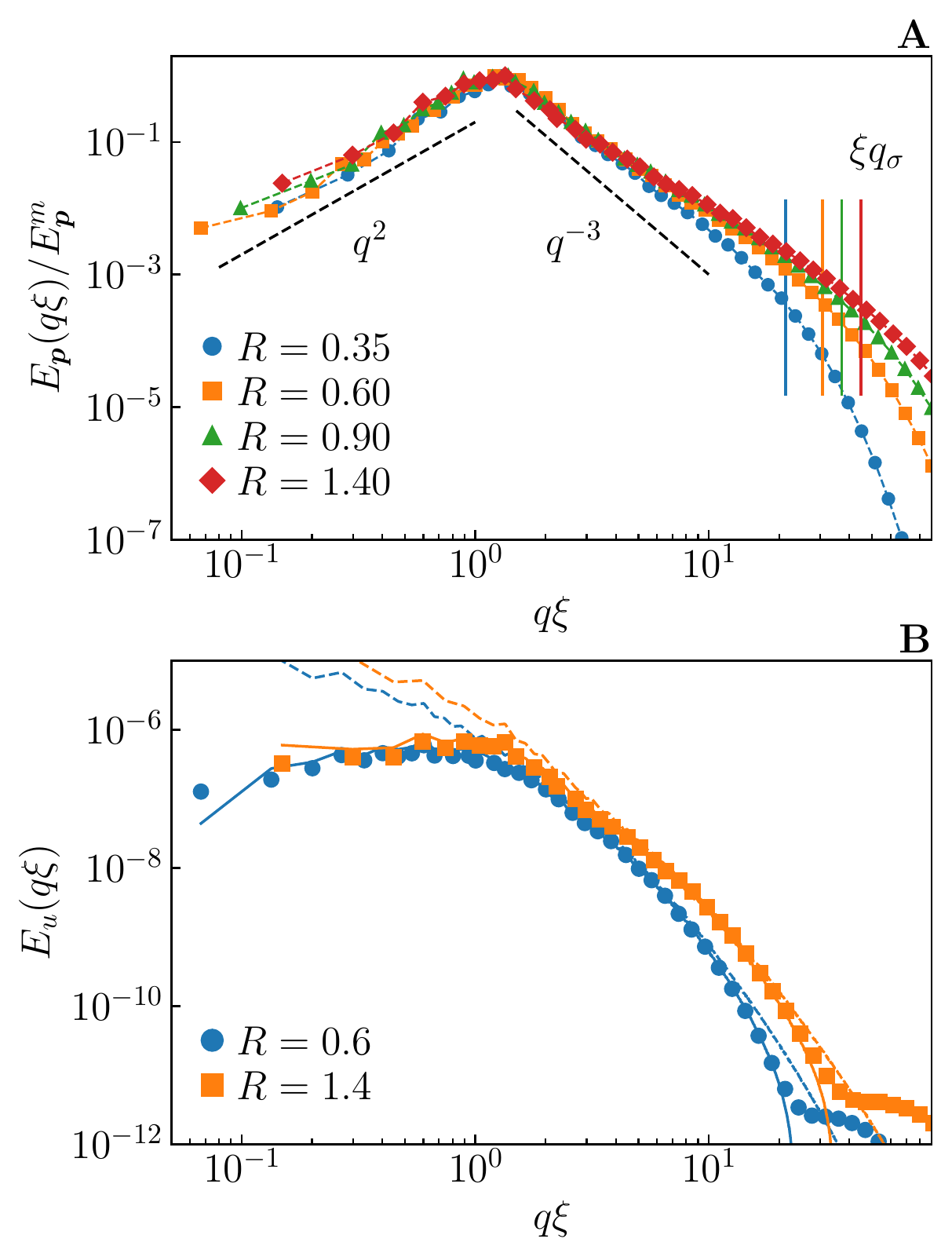}
    \caption{ \label{fig:spectra}
        (A) The order parameter energy spectrum $E_{\pol}(q\xi)$ for different values of R in both regime A and regime
        B. We scale the spectra by their respective peak values. $E_{\pol}(q\xi)$ shows a Porod scaling \cite{bray2002}
        for $1< q\xi < q_{\sigma}\xi$. (B) Comparison of the kinetic energy spectrum $E_{\vel}(q\xi)$ (solid markers) with
        \eqref{eq:visact} (solid lines) and \eqref{eq:visactsim} (dashed lines) for two different values of $R$.
    }
\end{figure}

The shell-averaged energy spectra for the velocity and the order parameter field is defined as
\begin{equation}\label{eq:spectra}\begin{aligned}
    E_{\vel}(q) &= \sum_{q-\frac{1}{2} \leq |\bm{m}| < q+\frac{1}{2}} |\widehat{\vel}_{\bm{m}}|^{2}, \mathrm{and}\\
    E_{\pol}(q) &= \sum_{q-\frac{1}{2} \leq |\bm{m}| < q+\frac{1}{2}} |\widehat{\pol}_{\bm{m}}|^{2},
\end{aligned}\end{equation}
where $\widehat{\vel}_{\bm{m}}$ and $\widehat{\pol}_{\bm{m}}$ are the Fourier coefficients of the velocity $\vel$ and
the order parameter $\pol$ fields. In \cref{fig:spectra}, we plot $E_{\vel}(q\xi)$ and $E_{\pol}(q\xi)$ for different
values of $R$. Consistent with our correlation function plots, we find that the spectra collapses onto single curve for
$q < q_\sigma$, where $q_\sigma=2\pi/\ell_\sigma$ with $\ell_\sigma\equiv \mu/\sqrt{\rho \sigma_0}$. Order parameter
spectra shows two distinct power law scaling regimes:
\begin{equation}\begin{aligned}
    E_{\pol}(q\xi) =
    \begin{cases}
        q^{2} \quad &\text{for}\quad \frac{2\pi}{L} < q < \frac{2\pi}{\xi}\\
        q^{-3} \quad &\text{for}\quad \frac{2\pi}{\xi} < q < \frac{2\pi}{\lsigma} ,
    \end{cases}
\end{aligned}\end{equation}
where $\lsigma = \mu/\sqrt{\rho\sigma_{0}}$, and $L$ determines the smallest wavenumber available in the simulation
domain. The $q^{-3}$ scaling is consistent with Porod's law prediction \cite{bray2002}.

The kinetic energy spectrum $E_{\vel}(q\xi)$ is determined by the balance of the viscous and active stresses
\begin{equation}\label{eq:visact}\begin{aligned}
    \llangle |\fft{\vel}{q}|^{2} \rrangle \approx -\frac{\so}{2\mu q^{2}} \llangle \fft{\vel}{q}^{*}\cdot\fft{\bm{f}}{q} \rrangle,
\end{aligned}\end{equation}
where $\fft{\bm{f}}{q} = \bm{\mathcal{P}}\cdot\left(i\bm{q}\cdot\fft{\pol\pol}{q}\right)$, $\bm{\mathcal{P}}$ is the
projection operator, and $\llangle \cdot \rrangle$ denotes temporal averaging.

In \cref{fig:spectra} we show an excellent agreement between the energy spectrum obtained directly from the velocity
field and using \eqref{eq:visact}, confirming the dominant balance between viscous and active stresses. Further assuming
$\fft{\bm{f}}{q}$ to be Gaussian random variables and using Gaussian integration by parts \cite{frisch1995} we get
\begin{equation}\label{eq:visactsim}\begin{aligned}
    \llangle |\fft{\vel}{q}|^{2} \rrangle
    = \left \langle \Bigg \langle \frac{\bm{\delta}\fft{\vel}{q}}{\bm{\delta}\fft{\bm{f}}{q}} \right \rangle \Bigg \rangle
    \llangle \fft{\bm{f}}{q}^{*}\cdot\fft{\bm{f}}{q}\rrangle
    \approx \frac{\so^{2}}{2\mu^{2} q^{4}}
    \llangle\fft{\bm{f}}{q}^{*}\cdot\fft{\bm{f}}{q}\rrangle.
\end{aligned}\end{equation}

We observe that the prediction \eqref{eq:visactsim} matches well with the energy spectrum for $1 <q \xi < q_\sigma \xi$
(see \cref{fig:spectra}).

\section{\label{sec:conclusions} Conclusions}

We study spatio-temporal properties of dense wet suspensions of polar active particles. Using a linear stability
analysis, we show how inertia can stabilize the orientational order against small perturbations. Our study reveals that
the dimensionless parameter $R\equiv \rho v_0^2/2\sigma_0$ characterizes the stability of the aligned state. Although
the neutral stability curve for DSPAP is identical to the previously studied case where concentration fluctuations are
rendered fast \cite{chatterjee2021}, our numerical studies reveal that their steady-state properties are very different. For
$R<R_2$ we observe defect turbulence, the order parameter flow consists of topological defects (vortices and saddles)
with no global polar or nematic order. We unravel a hidden defect-ordering transition by investigating the spatial
organization of defect centers. For $R=0$, defects are uniformly distributed and start to cluster with increasing $R$.
The clustering saturates around $R=R_1$ where we observe that the defects organize onto nearly one-dimensional
string-like structures. Finally, we show that the spectrum of the order-parameter field shows a Porod's scaling for
$q\xi>>1$ and a balance of viscous and apolar active stress determines the kinetic energy spectrum of the suspension
velocity.

S. R. acknowledges research support from a J. C. Bose Fellowship of the SERB, India.

\begin{figure*}
    \centering
    \includegraphics[width=\textwidth]{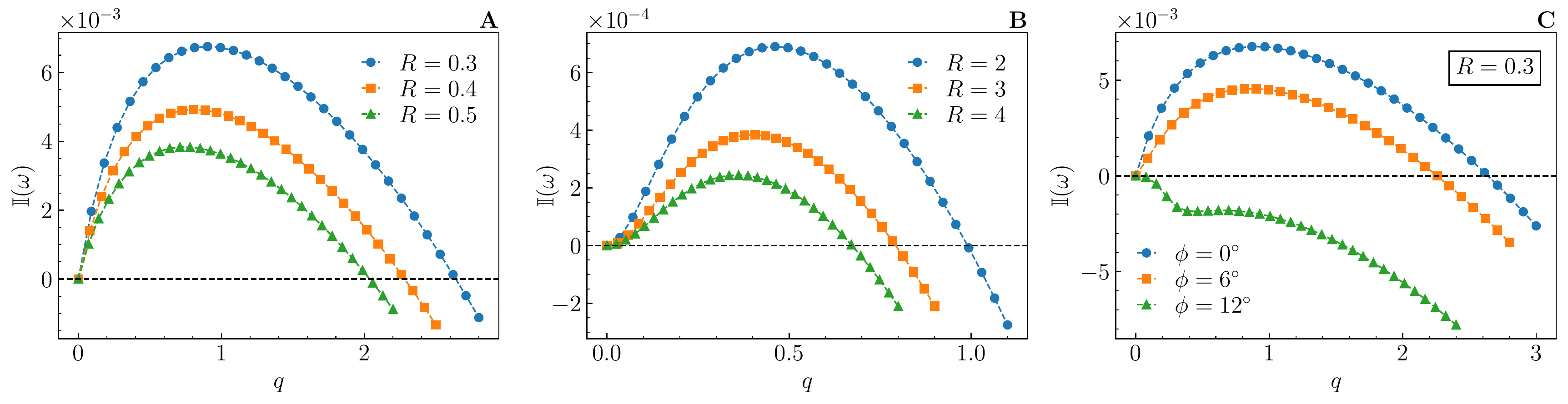}
    \caption{ \label{fig:growth_rate}
        (A, B) Growth rate $\imw$ vs. $q$ for the most unstable pure bend modes at various $R$ in regime A and B,
        respectively. With increasing $R$, the growth rate and the range of unstable wavenumbers decreases. (C) Growth
        rate $\imw$ vs. $q$ for the splay-bend modes at small values of $\phi$ for a fixed $R=0.3$.
    }
\end{figure*}
\appendix
\onecolumngrid
\section{\label{sec:appendix} Linear stability analysis}
We analyse the stability of the ordered state $\vel = 0,~\pol = \icap$ to small perturbations
$\delta\vel\equiv(\dux,\dun)$ and $\delta\pol\equiv(\dpx,\dpn)$, where $\ncap$ denotes the plane perpendicular to the
ordering. The linearised equations are
\begin{equation}\begin{aligned}
    \rho\dti\dux &= -\dx P +\mu\nabla^2\dux - \dx\left( \so\dpx +\lp K\nabla^2\dpx - 2\lambda\dpx\right), \\
    \rho\dti\dun &= -\dn P +\mu\nabla^2\dun - \dx\left(\so\dpn +\lp K\nabla^2\dpn\right), \\
    \dti\dpx &= -\dx\Pi -\vo\dx\dpx + \Gamma K\nabla^2\dpx - 2 \Gamma\dpx +\lambda\dx\dux, \\
    \dti\dpn &= -\dn\Pi -\vo\dx\dpn + \Gamma K\nabla^2\dpn +\lp\dx\dun +\lm\dn\dux. \\
\end{aligned}\end{equation}
Note that the perturbations also satisfy the incompressibility criteria, $\nabla\cdot\delta\vel = 0$, and
$\nabla\cdot\delta\pol = 0$. To proceed further, we consider monochromatic perturbations of the form $(\bm{\delta u},~
\bm{\delta p}) = (\widehat{\vel},~\widehat{\pol}) e^{i(\bm{q}\cdot\bm{x} -\omega t)}$,
where $\bm{q} \equiv \left(\qx \icap + \qn\cdot\ncap\right) = q \left(\cos\phi \icap + \sin\phi\ncap\right)$ is the
perturbation wavevector and $\omega = \mathbb{R}(\omega) + i\mathbb{I}(\omega)$. Using incompressibility constraint we
can eliminate pressure and longitudinal components to obtain
\begin{equation}\label{eq:linear}\begin{aligned}
    (-i\rho\omega+\mu q^2)\ukn &= 2i\lambda\qx\frac{\qn\cdot\pkn}{q^2}\qn-i\qx\left(\so-\lp Kq^2\right)\pkn\\
    (-i\omega+\Gamma Kq^2+i\vo\qx)\pkn &=-2\Gamma \frac{\qn\cdot\pkn}{q^2}\qn+i\lp\qx\ukn.
\end{aligned}\end{equation}
The linear system \eqref{eq:linear} is a set of four (two) coupled equations in three (two) dimensions and is easily
solved by decomposing into splay-bend and twist-bend modes \cite{simha2002, chatterjee2021, chandrasekhar1992}. Taking a
dot product with $\qn$ and solving for $\omega$ gives the following dispersion relation for the splay-bend modes
\begin{equation}\label{eq:splay-bend}\begin{aligned}
    2\omega^{s}_{\pm}
                      &= \vo q\ct - i\frac{\mup}{\rho} q^2 -2i\Gamma\sst \\
                      &\pm \frac{1}{\rho}\sqrt{(\rho\vo q\ct + i\mum q^2-2i\rho\Gamma\sst)^{2}
                      -4\rho\lp q^2 \cct (\so-\lp K q^2-2\lambda\sst)},
\end{aligned} \end{equation}
where $\mu_{\pm} = \mu (1 \pm \beta)$. Similarly, a cross product with $\qn$ yields for the three dimensional twist-bend
modes
\begin{equation}\label{eq:twist-bend}\begin{aligned}
    2\omega^{t}_{\pm} &= \vo q\ct - i\frac{\mup}{\rho} q^2
    \pm\frac{1}{\rho}\sqrt{(\rho\vo q\ct + i\mum q^2)^{2}-4\rho\lp q^2\cct(\so-\lp K q^2)}. \\
\end{aligned} \end{equation}
In two dimensions $\omega$ just has two solutions given by \eqref{eq:splay-bend}.  Additionally, in a general
description of polar suspensions, the concentration fluctuations only couples to the splay-bend modes, thus the
dispersion relation for twist-bend modes are identical for suspensions with fast \cite{chatterjee2021}, slow or no
concentration fluctuations.

\emph{Extensile Suspensions --} For extensile systems, the most unstable modes are the pure bend modes with $\phi=0$. In
this case, \eqref{eq:splay-bend} and \eqref{eq:twist-bend} are identical and as discussed in the main text, a small $q$
expansion reveals that the stability of the pure bend modes is governed by $R$. In \cref{fig:growth_rate}(A,B) we plot
the growth rate $\imw$ vs. $q$ for pure bend modes at different values of $R$ in regime A and B. The growth rate
for a given $q \neq 0$ decreases with increasing $R$, and is order of magnitude smaller in regime B as compared to
regime A.

The incompressibility constraint has major consequences on the stability of the $\phi \neq 0$ modes. The longitudinal
and transverse fluctuations are coupled to each other and thus $\dpx$ cannot be rendered \emph{fast}, as was done
previously in \cite{chatterjee2021}. Further, incompressibility constraint eliminates the splay deformations by an equal
and opposite contribution in the transverse direction which has a $q$-independent stabilizing effect. The relaxation
rate does not vanish in the $q\rightarrow0$ limit and at $q=0$ we have one non vanishing eigenvalue $\omega^{s}_{-} =
-2i\Gamma\sst$, which is a remnant of the coupling between $\dpx$ and $\dpn$. For small but nonzero $\phi$, the splay
contribution to these modes is small, and they are unstable in a manner similar to the pure bend modes but with a
smaller growth rate and a smaller range of unstable wavenumbers. In \cref{fig:growth_rate}(C) we plot $\imw$ vs. $q$ for
various values of $\phi$ at $R = 0.3$ and verify that indeed it is the case. The stability of the twist-bend modes is
identical to that of the pure-bend modes, i.e., depending on the various values of $R$, we obtain three distinct regimes
(See \cref{fig:phase-diagram} in main text), but with a $\phi-$dependent $R_1$ and $R_2$. The dispersion relation for
pure twist modes ($\phi=\pi/2$ in  \eqref{eq:twist-bend})  reduces to $\omega^{t}_{\pm} = -\frac{i}{2\rho}\left(\mup \mp
\mum \right) q^{2}$, implying that the pure twist modes are stable to linear perturbations.

\emph{Contractile Suspensions --} Contractile suspensions go unstable via two dimensional splay perturbations and as a
direct consequence of the incompressibility constraint on the order parameter, are always stable. This can also be
verified directly from \eqref{eq:splay-bend} and \eqref{eq:twist-bend}. For example, the pure splay modes $(\phi=\pi/2)$
relax with rates $\omega^{s} = \left(- i \frac{\mu}{\rho} q^2, -2 i \Gamma \right),$ and twist-bend modes are always
stable as $R < 0$.

\bibliography{Bibliography}
\bibliographystyle{apsrev4-2}
\end{document}

%% file: abbreviations.tex
\newcommand{\dd}{\mathop{}\!\mathrm{d}}

\newcommand{\postime}{(\bm{x},t)}
\newcommand{\icap}{\widehat{{x}}}
\newcommand{\jcap}{\widehat{{y}}}

\newcommand{\ncap}{\bm{\perp}}

\newcommand{\dx}{\partial_{x}}

\newcommand{\dti}{\partial_{t}}

\newcommand{\dn}{\nabla_{\ncap}}

\newcommand{\qx}{q_{x}}

\newcommand{\qn}{\bm{q_{\ncap}}}

\newcommand{\ct}{\cos\phi}
\newcommand{\sst}{\sin^2\phi}
\newcommand{\cct}{\cos^2\phi}

\newcommand{\vel}{\bm{u}}

\newcommand{\dux}{\delta u_{x}}

\newcommand{\dun}{\bm{\delta u_{\ncap}}}

\newcommand{\ukn}{\bm{\widehat{u}_{\ncap}}}

\newcommand{\pol}{\bm{p}}

\newcommand{\dpx}{\delta p_{x}}

\newcommand{\dpn}{\bm{\delta p_{\ncap}}}

\newcommand{\pkn}{\bm{\widehat{p}_{\ncap}}}

\newcommand{\res}{\bm{h}}

\newcommand{\lp}{\lambda_{\bm{+}}}
\newcommand{\lm}{\lambda_{\bm{-}}}
\newcommand{\vo}{v_{0}}

\newcommand{\so}{\sigma_{0}}
\newcommand{\mup}{\mu_{+}}
\newcommand{\mum}{\mu_{-}}
\newcommand{\Rdef}{\rho\vo^2/2\so}

\newcommand{\fft}[2]{\widehat{#1}_{\bm{#2}}}

\newcommand{\imw}{\mathbb{I}(\omega)}
\newcommand{\lsigma}{\ell_{\sigma}}

\newcommand{\llangle}{\langle\langle}%
\newcommand{\rrangle}{\rangle\rangle}%